# How We Lost the Internet

The Information Superhighway was paved with good intentions

Many of the problems experienced by developers and users of the current ICT environment are attributable to the definition of the Internet as a stateless communication stovepipe.

Micah D. Beck

Dept. of Electrical Engineering and Computer Science, University of Tennessee, Knoxville, mbeck@utk.edu

Terry R. Moore

Innovative Computing Laboratory, University of Tennessee, Knoxville, tmoore@icl.utk.edu

In this paper we reexamine an assumption that underpinned the development of the Internet architecture, namely that a *stateless and loosely synchronous point-to-point* datagram delivery service would be sufficient to meet the needs of all network applications, including those which deliver content and services to a mass audience at global scale. Such applications are inherently asynchronous and point-to-multipoint in nature. We explain how the inability of distributed systems based on this stateless datagram service to provide adequate and affordable support for them *within the public (I.e., universally shared and available) network* led to the development of *private overlay infrastructures*, specifically Content Delivery Networks and distributed Cloud data centers. We argue that the burdens imposed by reliance on these private overlays may have been an obstacle to achieving the Open Data Networking goals of early Internet advocates. The contradiction between those initial goals and the exploitative commercial imperatives of hypergiant overlay operators is offered as a possibly important reason for the negative impact of their most profitable applications (e.g., social media) and monetization strategies (e.g., targeted advertisement). We propose that one important step in resolving this contradiction may be to reconsider the adequacy Internet's stateless datagram service model.

CCS CONCEPTS • Networks Network architectures ~ Network design principles ~ Layering • Networks ~ Network services ~ Cloud computing • Networks ~ Network ~ In-network processing • Networks ~ Network types ~ Overlay and other logical network structures

**Additional Keywords and Phrases:** Internet history, stateful networking, multicast, content delivery, asynchronous communication

## 1 INTRODUCTION

The state of the Information and Communication Technology (ICT) environment— "the Internet," "the Web," and "the Cloud"—has been the subject of growing waves of complaint, outrage, and distress for more than a decade. The problems that typically provoke these sentiments are not without precedent. For example, businesses in the last century routinely used mass media as a conduit for marketing propaganda and tried to create monolithic retail mechanisms to capture and manipulate customers. But the spread of Internet-powered social media and of targeted advertising enhanced by consumer surveillance has refined and enhanced these strategies. This has made them much more worrisome and infuriating. Since the ICT hypergiants (e.g,. Google, Amazon, Meta) are leading players in the AI revolution, it seems unlikely that these trends will abate anytime soon.

For many commentators who lived through the development and growth of Internet-connected distributed systems, there is a stark dichotomy between the intentions and goals of the early champions of the Internet as the foundation for an "Open Data Network (ODN)" [1; 2] and the disturbing aspects of the ICT landscape we live in today. Explanations of this seeming paradox vary. In this paper we present a fundamental technical piece of the puzzle. In particular, we argue that the attempt to realize the global ODN vision of the 1990's by building on the Internet's end-to-end paradigm of communication infrastructure service exposed an essential limitation of reliance on stateless datagram delivery: *the lack of efficient in-network support for asynchronous point-to-multipoint services[1]*. This limitaiton created a need for auxiliary mechanisms outside the public *(I.e., universally shared and available)* network to support such services, which in turn created a ripe opportunity for the twenty-first century's exploitative business models.

Analyses of the consequences of these business models typically point to the obsession with constant growth of a company's user base, industry consolidation, and the concomitant emergence of monopoly power as the key factors in producing these unhappy results [3]. Such socio-economic factors are certainly real and important. However, in this paper we are concerned with the role of the early Internet's lack of efficient in-network support for asynchronous point-to-multipoint services. Our claim is that the absence of this functionality in the *public* Internet led to the growth of *private* infrastructure, which could and did become the tool of more rapacious and destructive forces. The complexity and expense of this private infrastructure required the providers of distributed, asynchronous services to develop sufficiently remunerative business models to pay for it. Such infrastructure was designed to cultivate and serve those markets that could support such business models. Being private, these providers were able to build businesses around tools and services that are more opaque and more invasive than those which are subject to community approval (e.g., by the Internet Engineering Task Force or the World Wide Web Consortium). The corporate consolidation and oligopolistic power that subsequently ensued raise complex issues that reach well beyond the domain of Computer Science and Engineering. However we argue that our community must take responsibility for reconsidering and, if possible redressing, the underlying technical factors that made such intrusive ICT business models both possible and to some extent unavoidable.

## 2  A TALE OF TWO "INTERNETS"

We begin by addressing a terminological difficulty in discussing the history and current state of the ICT environment. What the term "the Internet" denotes has changed over time, leading to a misunderstanding about the nature of the infrastructure that supports modern applications (e.g., Web sites) at scale. In order to avoid such confusion, we have assigned different names — Internet* and Internet++— to these distinct referents:

- **Internet\*** — We call the communication network that embodies most closely the original Internet architecture *the Internet\**. Also referred to as the Internet Protocol Suite, its most well-known elements are IP (the Internet Protocol) and TCP (the Transmission Control Protocol).[2] Internet* services comprise Layers 3 (Network) and 4 (Transport) of the IP Stack [4]. The asterisk is a reminder to the reader that the Internet* is not Internet++. The Internet* paradigm assumes that the stateless point-to-point datagram service provided by IP is an adequate basis for distributed systems that efficiently support asynchronous point-to-multipoint applications at scale.
- **Internet++** — We refer to the environment of applications and other services that are used today by billions of users and businesses worldwide as Internet++. This includes foundational services such as remote login and email, as well as ubiquitous facilities such as the World Wide Web, Web search, and social media. It also includes the additional infrastructures (e.g., Content Delivery Networks and distributed Cloud data centers) that have grown up to support today's more general application requirements.

---

[1] In the theory of distributed systems, the term "synchronous communication" must have a fixed upper bound on the delivery of any message and "asynchronous communication" has no upper bound. In applied networking, "synchronous communication" is understood to have a high probability of being delivered within a fixed time and "asynchronous communication" occurs on the time scale of storage systems.

[2] Other components of the Internet* include the Domain Name Service (DNS) and many that are not visible to end users include ICMP (the Internet Control Management Protocol) and internal and external (global) routing protocols such as OSPF (Open Shortest Path First), RIP (Routing Information Protocol), and BGP (Border Gateway Protocol).

The developers and early advocates of the Internet* argued for its adoption as the basis for America's National Information Infrastructure (NII), and by extension for the whole world's information infrastructure. Their vision of the future was something quite different from what Internet++ has become. That vision was an *Open Data Network* (ODN) that achieved four different goals (from the 1994 National Research Council report "Realizing the Information Future" [1], p. 44; see also [2], p. 36, 42): It was to be *open to users*, *open to service providers*, *open to network providers*, and *open to change*. But adopting the Internet* as the foundation for an ODN has instead produced the current Internet++, a global information infrastructure implemented as an *overlay* on the Internet* that has properties destructive to some of the core values informing this ODN vision, which many still hold dear. Table 1 summarizes the apparently paradoxical difference between the ODN that the NII aimed at, and the global overlay Internet++ ICT environment that has come to pass.

| Table 1: ODN goals for Internet* vs. Internet++ realities ||
| --- | --- |
| **The Open Data Network was supposed to be …** | **The Internet++ overlay data network …** |
| *Open to users*, permitting "universal connectivity as does the telephone system"; | *Effectively excludes large segments of the population from basic services* by requiring that such services include support for real time broadband interactivity. [5] |
| *Open to service providers*, delivering an open and accessible environment for competing commercial or intellectual interests. | *Encourages concentrations of power among both service and network providers* by requiring the deployment and connection of large numbers of replica servers throughout the network. This path has contributed to the emergence of a cartel of Internet++ "hypergiants" that use engineering dominance and overbearing market power to concentrate control over application provisioning. |
| *Open to network providers*, enabling any qualified network provider to attach and become a part of the aggregate of interconnected networks. | |
| *Open to change*, permitting "the introduction of new applications and services over time", as well as "new transmission, switching, and control technologies…." | *Results in network "ossification" (Sec. 3.1)* by excluding innovations other than those that can be implemented by routers supporting ultrahigh throughput and low latency. |

While a number of factors undoubtedly contributed to the eclipse of the ODN vision by Internet++ we see today, our hypothesis is that the assumed sufficiency of the Internet*'s stateless point-to-point datagram service set the stage for these paradoxical outcomes. Admittedly the considerations underlying our hypothesis are not purely technical. They also draw on historical records of the intentions and expectations of early Internet* advocates and formal principles of system design. As such we do not claim that the argument presented here authoritatively proves our hypothesis, nor that it can predict how the ICT environment might have evolved had different strategies been pursued. But we argue that the evidence is sufficient to show that, in the ongoing quest of the network research community for a way forward, possible alternatives to the stovepiped spanning layer of the Internet* should to more fully considered.

**2.1 The Internet Stovepipe**

In system architecture the term "stovepipe" refers to a collection of services implemented within each level of a layered system architecture. The services which comprise the stovepipe communicate only with other elements of that collection to export a service interface that is restricted to a specific scope. A stovepipe thus defines a vertical information conduit or "slice" of the layered system (Figure 1).

Accounts of ICT infrastructure architecture identify three fundamental stovepipes, namely Storage, Networking and Processing ([6], p. 215). Since the applications that these systems support necessarily combine the use of all three of these physical resources, the fact that they have to repeatedly access three separate interfaces atop three separate stovepipes (or "stacks") can be a source of inflexibility and inefficiency [7; 8].

### 2.2 The Internet* Stovepiped Spanning Layer

In a layered system, a *spanning layer* is a distinguished set of system elements that partitions the system's software stack vertically, with *applications* that are created using this interface above it, and the *lower-level services* required to support it below [9]. The purpose of a spanning layer is to enable *interoperability* among application implementations, allowing them to use different supporting services without any change to application code. Such interoperability is best served if the layer separating applications from the services that support them is *strict*, meaning that there is no direct access by applications to those services. A spanning layer is *stovepiped* if it offers to applications only a service that addresses a narrow/restricted type of functionality, *but which is insufficient to meet all the requirements of the necessary applications*. For example, like every digital application or service, an Internet* router must make internal use of data transfer, persistence, and processing resources. But the latter two primitive services are *encapsulated* within the implementation of the Internet Protocol Suite, which exports only a stateless datagram delivery service.[3]

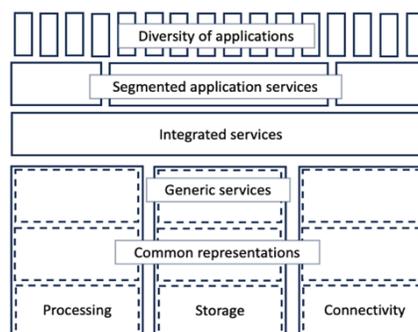

Figure 1: The Processing, Storage and Connectivity stovepipes supporting the distributed application services stack. Adapted from Messerschmidt and Szyperski. [6]

A stovepiped communication spanning layer, such as the Internetworking layer of the Internet* stack, does not support application requirements for distributed storage and processing. This may provoke two possible responses: The *first option* is to violate the strictness of the spanning layer, reaching beneath it to directly access underlying services that are supposed to be encapsulated behind its specialized interface. This can result in a system that does not have full interoperability, which can be inconvenient but, if limited, can still be managed. In fact, the ability for an implementation to violate the strictness of the spanning layer is sometimes even held up as a virtue [10]. The *second option* is to construct *non-shared* auxiliary infrastructure, i.e., an infrastructure not constrained by the spanning layer. Freed from spanning layer constraints, these auxiliaries can provide the resources and services that important classes of applications require. However, as the emergence of Internet++ shows, the necessary level of investment in such auxiliary infrastructure, and the increased economic power of its owners, can grow to overshadow the stovepiped infrastructure. In the case of Internet++, the outcome has been starkly at odds with the intentions and expectations of the Internet*'s original advocates.

### 2.3 Asynchronous Point-To-Multipoint Applications at Scale

There is a fundamental disconnect between the requirements of many of the most popular and profitable Internet++ applications (e.g., mass media and public digital utilities) and the nature of the network service provided by the Internet*. Many of the former require *asynchronous point-to-multipoint data and service delivery*, while the only ubiquitously deployed communication mechanism of Internet* is *a loosely synchronous one based on stateless point-to-point datagram*

---

[3] As Messerschmitt and Szypersky note, "The internet protocol can be viewed as a spanning layer, although it is limited to the connectivity stovepipe."

*delivery.*

**Distributed unicast drives Content Delivery Network infrastructure growth**

Consider a computer network that connects n clients through a graph of intermediate nodes connected by communication links. If the data is stored on a single server, the network traffic generated by iterated unicast delivery (Figure 2a) to *n* receivers is proportional to *n*p* where *p* is the average length of a path from source to receiver.

Multicast communication (implemented natively or in overlay) in a best-effort packet network proceeds through the forwarding of packets generated by a sender through a tree of forwarding nodes to reach a set of receivers at its leaves (Figure 2b). This form of point-to-multipoint communication is said to be "loosely synchronous." Because a sent packet is forwarded exactly once over each link in the multicast tree, the total network traffic generated by a single packet (measured in the number of links traversed in reaching receivers) is equal to the total number of such links[4]. In a perfectly balanced multicast forwarding tree, the number of such links is less than 2*n* (in a moderately unbalanced tree the coefficient of proportionality is higher). The traffic through a hierarchical arrangement of caches also generates traffic proportional to the number of links between caches.

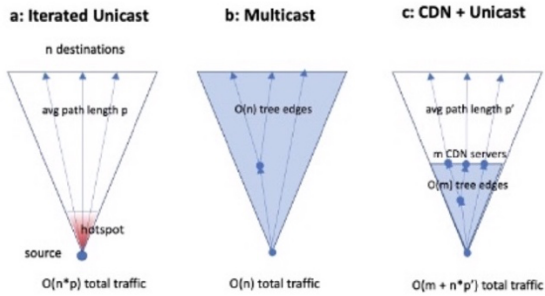

Figure 2: Unicast, Multicast and CDN

In a Content Delivery Network (CDN), data is uploaded to *m* intermediate storage-equipped servers (Figure 2c) If implemented efficiently, the traffic generated by the upload of data to intermediate servers will be proportional to the number of intermediate nodes.

Once uploaded to the CDN's servers, data is downloaded to each receiver by repeated use of unicast between the receiver and a server that is close to it in the network topology ("distributed unicast"). The total traffic generated during download from CDN servers is thus proportional to *n*p'*, where *p'* is the average distance between an endpoint and a close intermediate CDN server. Download may occur long after upload, so Content Delivery is said to be "asynchronous".

In CDN delivery the average load on any server is proportional to *(n*p')/m*. In fact, the ratio of annual global total traffic to server nodes deployed has stayed roughly constant at ~1.5 PB/server in recent years. Keeping up with the explosive growth in total Internet traffic (from 122 EB/mo. in 2017 to 293 EB/mo. in 2022) required an increase in the number of servers deployed by one of the largest global Content Delivery Networks (from 200K in 2017 to 350K in 2022). This kept the number of servers proportional to the total volume of data delivered to clients. These public industry statistics are not authoritative.

---

[4] Stated differently, multicast is a form of breadth first search, and thus generates traffic proportional to the number of links.

For example, the structure of simple file and object distribution applications is naturally asynchronous and point-to-multipoint: a file is published and advertised to users, who then request delivery of identical content at times and locations of their own choosing. To address this requirement using the point-to-point loosely synchronous functionality of Internet*, it has to be decomposed into two phases—"upload" and "download"—utilizing "iterated unicast" (see text box). By distinguishing between the point-to-multipoint structure of file delivery *as an application* and the use of an Internet* unicast *implementation*, one can see that the auxiliary server infrastructure that asynchronous point-to-multipoint applications require falls outside the shared network itself, and thus lies beyond the control of the Internet*'s implementers and advocates. At scale, the resources of the server and/or the network to which it connects tend toward exhaustion of local resources, resulting in inadequate responsiveness, inability to connect, or server instability—*the "hotspot" problem*. The necessity of responding to the hotspot problem was a primary driver of the development of Content Delivery Networks (CDNs), a fundamental element of Internet++ infrastructure.

The principles of the Internet* lead to a very different response to the hotspot problem than that which has been adopted in Internet++, which we discuss below. The canonical Internet* response is to advocate for the use of an efficient in-network solution, such as IP multicast, which delivers datagrams from a sender to multiple receivers using tree-structured forwarding. But despite decades of effort to promote its use, IP multicast has never been universally deployed [11]. Even if it were more widely used, IP multicast fails to meet the needs of asynchronous point-to-multipoint applications in a number of important respects: it is synchronous, it does not support congestion control or retransmission, and it is complex to implement and use [12]. These issues with IP multicast are not incidental to a particular protocol design; they are inherent to working within a successful point-to-point communication network which is growing explosively at a global scale. All of these issues could be addressed through the management of more persistent data state within the Internet* network. But as we discuss in Section 3.2, such strategies come into direct conflict with the "fast path" optimizations required for routers implementing unicast data forwarding at scale.

## 3 THE PARADOXICAL PATH FROM THE INTERNET* TO INTERNET++

A strong argument for using the Internet* as the foundation for a global Open Data Network was that its stateless model of point-to-point datagram delivery had been tested at reasonable scale. In 1994, the Internet* was a network of networks consisting "... of approximately 20,000 registered networks, some 2 million host computers, and 15 million users" [1]. As it turned out, stateless point-to-point datagram delivery could not continue to meet the application requirements and other objectives of the ODN vision as the system scaled up to serve everyone everywhere.

The explanation of this outcome depends on three key facts. First, the ability to provide asynchronous point-to-multipoint services is a fundamental requirement of multimedia streaming, Web services and other important applications. Second, asynchronous point-to-multipoint applications (as opposed to simple point-to-multipoint data delivery) often require nontrivial storage and processing at forwarding nodes. For example, a Web service may request the retrieval of data other than static objects. Per-response modifications to HTTP templates may range from the personalization of greetings and links to the complex presentation of individualized data. An HTTP server may make call-outs to local processes or remote servers. To implement such per-response processing within the network would require the use of a combination of point-to-multipoint data delivery, as well as in-network storage and processing.

Third, the efficient implementation of point-to-multipoint communication requires that routing be informed by network topology. Thus, all solutions that are conceived as "overlays" are eventually forced to use knowledge of network topology. They do so either inferentially, by positioning servers (e.g., caches and mirrors) close to the forwarding path between nominal source and destination, or directly, by using the Domain Name Service (DNS) or the Border Gateway Protocol (BGP) to pierce the encapsulation of network topology within the Internet* spanning layer. "Overlays" which pierce the Internet* spanning layer to access network topology act as extensions of that spanning layer.

Given these factors, as Internet* grew in the late '90s and its spectrum of applications diversified, the limitations of unicast datagram delivery had consequences both inside and outside the system. Internally, the effort to scale up distributed unicast to keep up with the growth of the network (e.g., number of endpoints, total traffic, etc.) created an imperative for the deployment of IP forwarding nodes that have the highest achievable capacity and sufficiently low latency. This ultimately led to reliance on the low latency "fastpath" which can prevent the adoption of efficient new in-network mechanisms in the Internet* spanning layer ("ossification")[5][11]. Externally, the need to provide asynchronous point-to-multipoint service led to the emergence of CDNs and distributed Cloud data centers, which included proprietary overlay routing mechanisms that repurposed the Internet*'s support infrastructure (e.g., DNS and BGP). Below we address each of these issues, and the way they frustrated the attempts of the network community to overcome the Internet*'s limitations.

### 3.1 How distributed unicast contributed to network ossification

The forwarding of a datagram arriving on the input port of an intermediate node to a single outbound port can be implemented using hardware acceleration—the "fast path"—which boosts peak performance to otherwise unattainable levels. Fast path acceleration allows the high levels of traffic generated by distributed unicast (see text box) to be handled by investment in server and router infrastructure, but it is a major contributor to network ossification.

Support for asynchronous point-to-multipoint or other new services are excluded because they increase forwarding delays on the routers due to services not being handled on the fast path. By 2006, Handley was already observing that "… using a new IP option would amount to a denial-of-service attack on the routing processor of many fast routers, and so such packets are highly likely to be filtered."[11] This progressive ossification has now reached the point at which "The Internet community has become inured to the public Internet*'s architectural stagnation and now greets talk of transformation with well-earned skepticism if not outright scorn." [13]

### 3.2 Why applications required another network (Internet++)

Given that innovation at the spanning layer was ruled out in practice by ossification, those aiming to provide asynchronous point-to-multipoint services, absent any regulations to constrain them, developed and relied on a sequence of work-arounds. Since these alternative measures were initially implemented in overlay using only application layer mechanisms, such work-arounds were once (and sometimes still are) viewed as applications of networking, rather than as alternative networks with their own architectures. These mechanisms use the components (subnetworks) which comprise the Internet* as building blocks and public facing on-ramps. But a little reflection makes it clear that they have evolved into full-fledged networks, implementing their own global topologies and policy mechanisms.

As early as the 1980's, the point-to-multipoint nature of file distribution was expressed through the use of "FTP mirroring". FTP mirror servers were not profit centers. Consequently, it was not foreseen that such servers would be constructed and operated specifically to implement large scale content delivery, or that they would become costly utilities that had to be supported by ample streams of income. Content Delivery Networks (CDNs) and then distributed Cloud networks have evolved to serve that purpose and, in so doing, have become new centers of power over Internet++.

One way to see that CDNs are full-fledged networks is by looking at the way they route the data that flows through them. In the Internet* architecture, the choice of a path for a datagram through the network is governed by routing protocols.

---

[5] While the requirement of low latency forwarding has been an important factor in network ossification, there are other contributing factors. These include the exhaustion of router resources such as memory bandwidth and processor capacity when datagrams are not handled on the fast path.

But antecedent to this routing of datagrams, a CDN makes a higher-level choice of which replica server a client request should be directed to. Making such choices to optimize network traffic and server load requires knowledge and management of the underlying resource topology, which makes it a kind of coarse-grained routing. CDNs reach below the spanning layer in order to access the underlying topology of the network. Many modern CDNs make the DNS the central mechanism for the direction of client requests to alternative servers [14]. Others leverage the "anycast" routing feature of BGP [15]. Thus, CDNs and distributed Clouds create their own networks by repurposing the components of the Internet*.

### 3.3 The impact of dependence on Internet++ on application business practices

The claim that dependence on Content Delivery Networks and distributed Cloud services reduces the *deployment scalability*[6] of applications, and that reliance on them has had a negative impact on business practices, requires some explanation. After all, these forms of private overlay infrastructure are deployed on a global scale, supporting countless Websites, streaming media, and business services. In addition, new Internet services have grown to support hundreds of millions (or, in the case of TikTok, billions) of users.

Internet++ overlay services combine Internet* transport with private networks and servers. Two important aspects of CDNs are that they are expensive to build and operate and that they do not provide equal access to the global community of Internet*-connected users. The cost of a CDN is passed on to their customers, who are the operators of Internet++ services and applications.

CDN and distributed Cloud operators do not generate income directly from subscriptions charged to end users, nor do they engage in aggressive business practices such as monetization through the sale of personal data. But dependence on them does impose the high costs that all but require such practices. This is simply a consequence of the lack of deployment scalability in the Internet++ overlay architecture.

The Internet* was designed to enable full connectivity between every pair of communicating endpoints. Internet++ overlay networks, on the other hand, can place content on specific servers and manipulate global routing to restrict service to specific regions of the world. This may be done to reduce costs, to enforce content licensing agreements, or to enforce governmental policies restricting the open exchange of information.[7]

Innumerable content and service providers, including many that are not-for-profit or which are designed to serve the public good, are deployed using Internet++. Any service that generates high levels of traffic must pay this cost. An example of success in innovation at global scale is TikTok, which serves 1.7 billion end users. TikTok is owned by Bytedance, a company valued at over $250 billion. It is reported to have used distributed infrastructure deployed worldwide by a number of companies, including Alibaba, AWS, and Oracle.

The addictive nature of TikTok's content selection algorithm, the degree to which it collects and monetizes user data, and the specter of its control by the People's Republic of China are major political and economic concerns in much of the world. TikTok has been banned from many governmental networks in the US, and there have been attempts to ban it outright in some states. TikTok's model of capital-intensive growth coupled with aggressive monetization of end user attention is very different from the vision of Open Data Networking. By contrast, as of 2023, the open federated social media platform Mastodon had under 10 million total users, with the number of active users (under 2 million), and a server deployment that has not grown during the year.

---

[6] Deployment scalability, discussed in more detail in [16], refers to the tendency of a service to be deployed widely and a to be voluntarily adopted as a de facto standard.
[7] The leveraging of Internet++ control over access is noted in a recent article: "While the Internet faces many challenges, we think that none are as important to its future as resolving the dichotomy of visions between the one that animated the early Internet community, which is that of a uniform Internet connecting all users and largely divorced from services offered at endpoints, and the one represented by the emerging large private Cloud and Content Provider networks, in which the use of in-network processing provides better performance but only to clients of their cloud or content services." [17]

## 4 THE TAIL WAGS THE DOG: PAYING THE BILL FOR ICT INFRASTRUCTURE

The Internet* architecture is nominally store-and-forward in design [18], and thus asynchronous. However, in practice it has always been used as a loosely synchronous network. A strongly asynchronous network (i.e., networks that have no bounds on latency or loss of individual packets) cannot necessarily support some demanding applications that were part of the original global ODN vision and are important to Internet++. These include real-time video streaming, telepresence (e.g., audiovisual conferencing), and remote control. The idea that the Internet* architecture could and should support such low latency applications implicitly assumes an implementation that is statistically synchronous, meaning that there are bounds on latency, jitter and loss over short periods of time, and on continuous service availability over long periods.

But a recent result in the theory of layered systems of infrastructure services—the Hourglass Theorem [16]—predicts that such strengthening comes at a non-trivial cost. The Hourglass Theorem describes an inherent tradeoff between the logical strength of a spanning layer, the collection of applications that it can support, and the variety of underlying services that can implement it. As a design principle, it tells us that adopting the *weakest* spanning layer that can support a specified collection of necessary applications tends to maximize *the variety* of underlying services that can support it. Any strengthening of the spanning layer beyond such "minimal sufficiency" will tend to diminish its wide and easy adoption.

Accordingly, requiring the network's spanning layer to have a statistically synchronous implementation in order to support applications which require very low latency will reduce the availability of network services to some users. In many cases those excluded users *could* make productive use (sufficient to *their needs*) of implementation strategies that provide weaker guarantees, including some that are far lower in cost or much easier to implement. As we have argued elsewhere [5], a low latency requirement for the spanning layer may be one important reason for the stubborn persistence of the digital divide. It may well be that the proportion of the connected population that can use applications requiring the highest quality synchronous connectivity may actually decline in the future as stronger service guarantees are adopted.

So far we have argued that Internet++ evolved as a private response to the inability of the network research community to evolve a universally shared and available Internet* to meet the need for asynchronous point-to-multipoint services and content distribution. Following a similar path, several major distributed Cloud data center networks also emerged from specific application domains. Cloud services defined a new paradigm for the creation of applications with huge user communities. The private nature of such CDN and distributed Cloud infrastructure obviated the need for that infrastructure to be usable by application providers who are not their paying customers. Moreover, private ownership and centralized administration of such essential infrastructure in turn presented an opportunity to maximize the profitability of online services by taking advantage of features that are not supported by public networks, which must serve everyone. In the case of Internet++, such features include the following: the ability to surveil the activity of users and to collect and resell their personal data; the ability to utilize that data to create a business model that promises much more accurately targeted and effective advertising; and the ability to maximize the profit such models generate by creating engagement-maximizing algorithms that increase the amount of time and attention users devote to online applications, such as social media and gaming.

It seemed plausible in 1994 to assume that a network based on a stateless model of unicast datagram delivery could support an Open Data Network that would efficiently deliver, at scale, asynchronous point-to-multipoint services and applications. But the stateless datagram sufficiency assumption has so far turned out to be false. What has proved true is that, by appropriating Internet* mechanisms for their own ends, the private overlay service providers of Internet++ have been able to overcome the limitations of the Internet* model, delivering services and applications envisioned for ODN, and more. However, as we suggested in Table 1 above, the price for this way of overcoming the shared public Internet*'s stateless point-to-point delivery model has been the sacrifice of ODN's defining goals.

## 5 FIGHTING THE GOOD FIGHT: OVERLAY AND UNDERLAY

We acknowledge that a significant segment of the Computer Science and Engineering community understood the inefficiency of iterated unicast and made extensive efforts to improve on it. Beyond attempts to deploy IP multicast, discussed above, the Active Networking community of the 1990s [19] sought to address the need for available data persistence and processing capabilities available within the network. The Web caching movement created an overlay form of URL-based content delivery of static objects [20]. Turning Web caches into "middleboxes" that can provide more general services has proved challenging, and their functionality is now largely limited to security firewalls [21]. Named Data Networking attempted to combine static object caching with a publish/subscribe service model to replace or augment the IP spanning layer [22].Advanced engineering testbeds such as the PlanetLab [23], the Generalized Environment for Network Innovation (GENI) [24], and FABRIC [25] have resulted in some incremental innovations to the current network architecture, such as Software Defined Networking [26] and Network Function Virtualization [27]. Limitations of length prevent us from a full enumeration of the persistent, determined and well-funded research efforts which have tried, without notable success, to supplant Internet++.

The intent of this article is not to claim that an alternative history was possible but to suggest where we should be looking for a path out of the current social and technological cul-de-sac. The situation we find ourselves in has been described as a stark paradox: the stovepiped IP spanning layer (i.e., Layer 3 of the Internet* protocol stack, or L3) has to be deployed on every network intermediate node, and the use of every network application depends on it. Consequently, it is *de facto* impossible to change. But if L3 impossible to change, then it cannot be extended to support applications that generate new application requirements. "This is why the Internet, which is otherwise so extensible at the higher and lower level layers, has a rigid and unchanging service model." [28]

This reference to the layers above and below the spanning layer suggests that there are at least two possible routes of escape from this paradox. On one hand, as already discussed above, the traditional approach is to work at layers *above* the spanning layer to create a network overlay. Some of today's thought leaders in network research have proposed a new version of the overlay strategy, advocating the creation of an Extensible Internet (EI). To accommodate the ossified nature of the Internet* at L3, EI would offer an overlay implemented primarily on new service nodes located at the edge of the public network's core. EI's seeks to provide the limited storage and processing resources required for the implementation of new services that could be closely regulated by the community [13].

Alternatively, looking *beneath* L3 seems to offer a different, if unorthodox, route to extensibility. Specifically, we might seek an "underlay" approach that moves the spanning layer of the public, wide area infrastructure to a *lower level*. This would entail the creation of a more general abstraction (i.e., one that includes storage and processing) of network nodes and local communication resources, which are traditionally referred to as Link Layer (L2) resources. The idea of locating the spanning layer at L2 was broached in a 1996 white paper in which David Clark discussed the potential innovations that might be unleashed through heterogeneity at L3, via a network stack that had the shape of "a funnel placed on its end" rather than an hourglass [9]. GENI engineers also considered the possibility of a spanning layer at L2 [24].

But in any scenario, whether overlay or underlay, the key challenge in the design of a new common spanning layer is deployment scalability. In that regard, the Hourglass Theorem [16] may prove to be an invaluable tool. For example, since the long-term goal of an overlay spanning layer (at "L3.5") is universal deployment across the public network, if the services it defines are logically stronger than those of the Internet*'s L3, the Hourglass Theorem suggests that universal access to this new level of infrastructure may impose burdens that some edge networks may be unable to meet. This, in turn, may stymie the EI goal of its universal deployment. Likewise, the goal of extreme deployment scalability will confront any proposed underlay solution with equally intimidating trade-off problems, which the Hourglass Theorem may also help infrastructure architects to analyze and address.

## 6 CONCLUSION: GETTING BEYOND THE INTERNET STOVEPIPE

We began our discussion of the Internet stovepipe by differentiating the Internet* from Internet++; and we proceeded to offer an explanation of how the technical limitations of the former, due largely to its stateless point-to-point datagram service model, paved the way to the latter. The result has been an application environment that continues to be called "the Internet" but which has largely abandoned the goals of Open Data Networking in favor of burdensome strategies that primarily serve well-financed commercial interests or highly subsidized governmental ones. We cannot rule out the possibility that, even if efficient asynchronous point-to-multipoint services had been universally deployed, other bottlenecks would not have emerged and impacted the Internet* environment in other ways. The pursuit of profit and market dominance might still have overwhelmed the altruistic or communitarian intentions of some of the Internet*'s early advocates. Yet examples such as email and Web browsing have shown that an open, scalable service architecture *can* support both interoperable base offerings and proprietary ones funded by subscription and/or advertising. Open competition does not rule out abusive business practices, but it leaves room for alternatives which can apply a moderating force.

While it is certainly more comfortable to blame corporate greed or governmental failures for the darker consequences of the Internet revolution, it is hard to see how the situation can be substantially improved until the Internet community takes responsibility for its acquiescence in a series of workarounds to the limitations of the Internet* architecture. These so-called "barnacles" [29] have turned out to be fatal to the goals of Open Data Networking. Path dependence due to sunk investment, whether in material infrastructure or cherished beliefs, can make a reconsideration of current practice seem impossible. And this reluctance is made even stronger when the theories that need to be questioned lie at the foundation of technologies and infrastructures that billions of people depend on daily. Such aversion to the reconsideration of fundamental assumptions is said to have led Max Planck to observe that science advances only one funeral at a time [30]. Today we submit to entrenched orthodoxy while cashing the checks intended for the creation of a more equitable and sustainable ICT environment. If we despair of solving the most troubling and vexing problems of infrastructure design, then no one else will save us from the consequences of our own past creations.

## 7 REFERENCES

[1] National Research Council. 1994 Realizing the Information Future. National Academies Press.
[2] Council, N. R. 2001 The Internet's Coming of Age. The National Academies Press.
[3] Kulwin, N. Apr 13, 2018. An Apology For The Internet—From The Architects Who Built It. New York Magazine.
[4] 1989. RFC1122: Requirements for Internet Hosts - Communication Layers. 10.17487/rfc1122.
[5] Beck, M. and Moore, T. 2022. Is Universal Broadband Service Impossible. 2022 IEEE 19th International Conference on Mobile Ad Hoc and Smart Systems, MASS, 403-409. DOI=10.1109/MASS56207.2022.00064.
[6] Messerschmitt, D. G. and Szyperski, C. 2003 Software ecosystem: understanding an indispensable technology and industry. MIT press Cambridge.
[7] National Research Council. 2001 Looking Over the Fence at Networks. National Academies Press.
[8] Peterson, L., Anderson, T., Culler, D., and Roscoe, T. 2003. A blueprint for introducing disruptive technology into the Internet. ACM SIGCOMM Computer Communication Review. 33, 1, 59-64. DOI=10.1145/774763.774772.
[9] Clark, D. D. 1995. Interoperation, open interfaces and protocol architecture. The Unpredictable Certainty: White Papers. 2, 133-144.
[10] Peterson, L. L. and Davie, B. S. 2011. Computer Networks: A Systems Approach.
[11] Handley, M. 2006. Why the Internet only just works. BT Technology Journal. 24, 3, 119-129.

[12] Diot, C., Levine, B. N., Lyles, B., Kassem, H., and Balensiefen, D. 2000. Deployment issues for the IP multicast service and architecture. IEEE network. 14, 1, 78-88.
[13] Balakrishnan, H., Banerjee, S., Cidon, I., Culler, D., Estrin, D., Katz-Bassett, E., Krishnamurthy, A., McCauley, M., McKeown, N., Panda, A., Ratnasamy, S., Rexford, J., Schapira, M., Shenker, S., Stoica, I., Tennenhouse, D., Vahdat, A., and Zegura, E. 2021. Revitalizing the public internet by making it extensible. ACM SIGCOMM Computer Communication Review. 51, 2, 18-24. DOI=10.1145/3464994.3464998.
[14] Wang, Z., Huang, J., and Rose, S. 2018. Evolution and challenges of DNS-based CDNs. Digit Commun Netw. 4, 4, 10.1016/j.dcan.2017.07.005.
[15] Sarat, S., Pappas, V., and Terzis, A. 2005. On the use of anycast in DNS. ACM SIGMETRICS Performance Evaluation Review. 33, 1, 394-395. DOI=10.1145/1071690.1064271.
[16] Beck, M. 2019. On the hourglass model. Communications of the ACM. 62, 7, 48-57. DOI=10.1145/3274770.
[17] Mccauley, J., Shenker, S., and Varghese, G. 2023. Extracting the Essential Simplicity of the Internet. Communications of the ACM. 66, 2, 64-74. DOI=10.1145/3547137.
[18] Clark, D. 1988. The design philosophy of the DARPA Internet protocols. Symposium proceedings on Communications architectures and protocols, 106-114.
[19] Tennenhouse, D. L., Smith, J. M., Sincoskie, W. D., Wetherall, D. J., and Minden, G. J. 1997. A survey of active network research. IEEE Communications Magazine. 35, 1, 80-86. DOI=10.1109/35.568214.
[20] Gadde, S., Chase, J., and Rabinovich, M. 2001. Web caching and content distribution: a view from the interior. Computer Communications. 24, 2, 222-231. DOI=10.1016/s0140-3664(00)00318-2.
[21] Walfish, M., Stribling, J., Krohn, M., Balakrishnan, H., Morris, R., and Shenker, S. 2004. Middleboxes no longer considered harmful. Proceedings of the 6th conference on Symposium on Operating Systems Design & Implementation - Volume 6, 15.
[22] Zhang, L., Afanasyev, A., Burke, J., Jacobson, V., claffy, K., Crowley, P., Papadopoulos, C., Wang, L., and Zhang, B. 2014. Named data networking. ACM SIGCOMM Computer Communication Review. 44, 3, 66-73. DOI=10.1145/2656877.2656887.
[23] Peterson, L. and Roscoe, T. 2006. The design principles of PlanetLab. ACM SIGOPS Operating Systems Review. 40, 1, 11-16. DOI=10.1145/1113361.1113367.
[24] 2016. The GENI Book. 10.1007/978-3-319-33769-2.
[25] Baldin, I., Nikolich, A., Griffioen, J., Monga, I. I. S., Wang, K.-C., Lehman, T., and Ruth, P. 2019. FABRIC: A National-Scale Programmable Experimental Network Infrastructure. IEEE Internet Computing. 23, 6, 38-47. DOI=10.1109/MIC.2019.2958545.
[26] Kreutz, D., Ramos, F. M. V., Veríssimo, P. E., Rothenberg, C. E., Azodolmolky, S., and Uhlig, S. 2015. Software-Defined Networking: A Comprehensive Survey. Proceedings of the IEEE. 103, 1, 14-76. DOI=10.1109/JPROC.2014.2371999.
[27] Mijumbi, R., Serrat, J., Gorricho, J.-L., Bouten, N., De Turck, F., and Boutaba, R. 2016. Network Function Virtualization: State-of-the-Art and Research Challenges. IEEE Communications Surveys & Tutorials. 18, 1, 236-262. DOI=10.1109/COMST.2015.2477041.
[28] Shenker, S. April 14, 2022. Creating an Extensible Internet. https://blog.apnic.net/2022/04/14/creating-an-extensible-internet/
[29] Anderson, T., Peterson, L., Shenker, S., and Tuner, J. 2005. Overcoming barriers to disruptive innovation in networking Report of NSF Workshop. 2005, 26.
[30] Wikipedia, C. 2023 Planck's principle --- Wikipedia, The Free Encyclopedia.